\documentclass[onecolumn,12pt]{article} 

\title{A link between the maximum entropy approach and the variational entropy form}
\author{E.V. Vakarin, J.P. Badiali\\
UMR 7575 LECIME ENSCP-UPMC,\\ 11 rue P. et M. Curie, 75231 Cedex
05, Paris, France}

\date{}
\begin{document}
\maketitle
\begin{abstract}
The maximum entropy approach operating with quite general entropy measure
and constraint is considered. It is demonstrated that for a conditional or parametrized probability distribution $f(x|\mu)$ there is a "universal" relation
among the entropy rate and the functions appearing in the constraint.
It is shown that the recently proposed variational formulation of the entropic functional can be obtained as a consequence of this relation, that is  from the maximum entropy principle. This resolves certain puzzling points appeared in the variational approach.  
\end{abstract}

\section{Introduction}
The maximum entropy approach\cite{Jaynes} is extensively invoked in the solution of inference problems appearing in various fields (from physics to biology). The approach allows one to calculate a probability distribution of relevant variables from an incomplete knowledge of the microscopic mechanisms governing the system evolution. The central idea is to find a least biased distribution consistent with the information at hand. Therefore, the scheme operates with two main ingredients: the entropy measure (that estimates the uncertainty associated with the probability distribution) and the constraints encoding the available information. In this context various entropy functionals are introduced (see \cite{next} for a recent review) in order to derive various non-exponential probability distributions observed in nature.
These are generalizations of the famous Shannon form (\ref{shannon}) that appears from the generalized forms as some limiting case. However, many of these generalized entropies are non-additive for statistically independent subsystems. This fact with its consequences is still a subject of extensive debates \cite{next}. 

In a situation when many generalized entropy forms are being proposed \cite{Thurner} it might seem that the
maximum entropy approach suffers from an ambiguity \cite{Plastino1}: "in a sense that any probability distribution seems to be derivable from the maximization of any entropic measure if an appropriate constraint is used". In this respect it has been argued \cite{Plastino1} that such an ambiguity does not appear when one searches for a parametric family of distributions
$f(x|\mu)$. Here $x$ is a relevant fluctuating variable and $\mu$ is an external parameter independent of $x$. For instance, such a parameter could enter through the mean value \cite{JCPcomp} or the distribution width varying in response to some external perturbation, like insertion into a host matrix \cite{condpol}.  
On the other hand, we have to note that we are not free to modify the constraints without logical necessity because they contain the information (or our theoretical model) on the processes running in the system of interest. For example, a $q-$exponential distribution can be easily obtained through maximizing the Tsallis entropy (\ref{tsallis}) with the constraint (\ref{overbar}) on the mean value of the relevant variable, but the
physics is hidden in the form of the entropy functional and the meaning of the index $q$ remains unclear. The same $q-$exponential distribution has been obtained within the superstatistical approach \cite{Beck} that operates with two subsystems fluctuating at different time scales.  
On the other hand, it has been demonstrated \cite{PREpow,AIP} that the same distribution can be obtained from the Shannon form (\ref{shannon}) with quite transparent
physical constraints encoding a departure from the standard equilibrium conditions. Namely,
a reservoir with fluctuating temperature in contact with a system whose phase space is restricted. This corresponds to a non-equilibrium stationary state and 
the magnitude of $q$ measures the distance from the equilibrium corresponding to the standard exponential (Gibbsian) distribution. A quite similar idea of non-equilibrium hyperensembles, leading to non-exponential distributions, has recently been proposed \cite{Crooks}. All these arguments demonstrate that a given distribution can appear as a result of various (not necessarily similar) processes. 

In this connection an inverse problem has been proposed \cite{Abe,Wang1,Wang2}. 
This variational scheme allows one to find the entropy functional that should be chosen in order to find a given distribution with a given constraint. It has been found surprising \cite{Wang1} that for various distributions the functionals derived in this approach coincide with the generalized entropy forms used in the maximum entropy
procedure.  

In this paper we consider the maximum entropy procedure operating with quite arbitrary entropy functional and the constraint imposed on the relevant variable $x$. This allows us to derive
a relation among the entropy rate, the observed function and the average of its model estimation. This result does not depend (explicitly) neither on the entropy form nor on the form of the constraint. Based on this relation we make a link between the maximum entropy approach and the variational scheme \cite{Abe,Wang1,Wang2}. This explains why the results of the direct (maximum entropy) and the inverse (variational) problems are coherent. 

\section{Maximum entropy approach}

In the light of what is discussed above, let us consider the entropy functional $H(\mu)$
of a quite general form
\begin{equation}
H(\mu)=D\left(\Phi\left[ f(x|\mu) \right]  \right)
\label{H}
\end{equation}    
Here $H$ is simultaneously a function of the parameter $\mu$ and a functional of the parametrized distribution $f(x|\mu)$, $D(t)$ is some function with appropriate behavior. 
The functional $\Phi\left[ f(x|\mu) \right]$ can be represented as 
\begin{equation}
\Phi\left[ f(x|\mu) \right]=\int{dx \varphi \left(f(x|\mu) \right)},
\label{Phi}
\end{equation}  
where $\varphi(p)$ is another function ensuring suitable overall properties (continuity, concavity,...). Combining eqs. (\ref{H}) and (\ref{Phi}) we arrive at
\begin{equation}
H(\mu)=D\left( \int{dx \varphi \left(f(x|\mu) \right)} \right)
\label{HH}
\end{equation}
It is clear that practically all generalized entropy functionals appearing in the current literature can be deduced from the definition above (except the forms involving the distribution derivatives like the Fisher form). For instance, for $D(t)=t$, $\varphi(p)=-p\ln p$ we recover the Shannon entropy (\ref{shannon}). The choice $D(t)=t$, $\varphi(p)=(p-p^q)/(q-1)$ corresponds to the Tsallis form (\ref{tsallis}). The Renyi form is recovered if $D(t)=\ln t$, $\varphi(p)=p^q/(1-q)$.

Let us consider a problem of maximizing $H(\mu)$ under two natural constraints - the normalization
\begin{equation}
\int{dx f(x|\mu)}=1
\label{norm}
\end{equation}
and some known average $\theta(\mu)$ of a conditional response function $\theta(\mu|x)$
\begin{equation}
\theta(\mu)={\overline{\theta(\mu|x)}}=\int{dx f(x|\mu)\theta(\mu|x)}.
\label{constraint}
\end{equation} 
Here the overbar denotes the corresponding average taken with the distribution
$f(x|\mu)$.
Usually the average $\theta(\mu)$ corresponds to the observed behavior (e.g. experimental data) and the conditional function $\theta(\mu|x)$ represents our theoretical estimation (a model) of this behavior for a given state $x$ of the random environment. The latter is characterized by a probability distribution  $f(x|\mu)$ depending on the "driving" parameter $\mu$ (for instance, this could be an external field, pressure or chemical potential).  

Thus we have to find a variation of the following Lagrangian
\begin{equation}
L=H(\mu)+\kappa\left[\int{dx f(x|\mu)\theta(\mu|x)}- \theta(\mu)\right]+
\lambda\left[\int{dx f(x|\mu)}-1 \right],
\label{lagrangian}
\end{equation}
where $\kappa$ and $\lambda$ are the Lagrange multipliers which should be found
from the constraints (\ref{norm}) and (\ref{constraint}). The variation procedure
$\delta L/\delta f(x|\mu)=0$
leads to
\begin{equation}
\frac{\partial D\left(\Phi\left[ f(x|\mu) \right]  \right)}
{\partial \Phi\left[ f(x|\mu) \right]}
\frac{\partial \varphi \left(f(x|\mu) \right)}
{\partial f(x|\mu)}=-\kappa \theta(\mu|x)-\lambda
\label{variation}
\end{equation}
Let us consider now the variation of $H(\mu)$ with respect to the parameter $\mu$. From eqs. (\ref{H}) and (\ref{Phi}) we obtain
\begin{equation}
\frac{\partial H(\mu)}{\partial \mu}=
\frac{\partial D\left(\Phi\left[ f(x|\mu) \right]  \right)}
{\partial \Phi\left[ f(x|\mu) \right]}
\frac{\partial \Phi\left[ f(x|\mu) \right]}
{\partial \mu};\qquad
\frac{\partial \Phi\left[ f(x|\mu) \right]}
{\partial \mu}=\int{dx \frac{\partial \varphi \left(f(x|\mu) \right)}
{\partial f(x|\mu)}}
\frac{\partial f(x|\mu)}{\partial \mu}
\end{equation} 
Here $\partial D\left(\Phi\left[ f(x|\mu) \right]  \right)/
\partial \Phi\left[ f(x|\mu) \right]$ could be a function of $\mu$. Anyway it does not depend on $x$ since $\Phi\left[ f(x|\mu) \right]$ is a functional of the probability distribution and simultaneously a function of $\mu$.
Therefore, combining these two equations we obtain
\begin{equation}
\frac{\partial H(\mu)}{\partial \mu}=
\int{dx 
\frac{\partial D\left(\Phi\left[ f(x|\mu) \right]  \right)}
{\partial \Phi\left[ f(x|\mu) \right]}
\frac{\partial \varphi \left(f(x|\mu) \right)}
{\partial f(x|\mu)}
\frac{\partial f(x|\mu)}{\partial \mu}
}
\end{equation}
Using now the condition (\ref{variation}) for the maximum entropy we arrive at
\begin{equation}
\frac{\partial H(\mu)}{\partial \mu}=
-\kappa\int{dx \theta(\mu|x)\frac{\partial f(x|\mu)}{\partial \mu}}
-\lambda\int{dx \frac{\partial f(x|\mu)}{\partial \mu}}
\end{equation}
It is clear that the last term disappears because of the normalization condition (\ref{norm}) and we finally get the maximum entropy rate as
\begin{equation}
\frac{\partial H(\mu)}{\partial \mu}=
-\kappa\int{dx \theta(\mu|x)\frac{\partial f(x|\mu)}{\partial \mu}}.
\label{vvar0}
\end{equation}
Now differentiating the constraint (\ref{constraint}) we obtain
\begin{equation}
\frac{\partial \theta (\mu)}{\partial \mu}=
\int{dx \frac{\partial \theta(\mu|x)}{\partial \mu}} f(x|\mu)+
\int{dx \theta(\mu|x) \frac{\partial f(x|\mu)}{\partial \mu}}
\end{equation} 
Combining these two equations we arrive at
\begin{equation}
\frac{\partial H(\mu)}{\partial \mu}=
-\kappa \left[
\frac{\partial \theta(\mu)}{\partial \mu}-
\overline{\frac{\partial \theta(\mu|x)}{\partial \mu}}
\right]
\label{dHdm}
\end{equation}  
Therefore, we have a "universal" relation among the entropy rate, the observed behavior and the average of the model estimation. This relation does not depend on the form of the entropic functional (the form (\ref{H}) is quite general) and thus could be helpful for drawing conclusions of general validity. A brief discussion of (\ref{dHdm}) for the case of the Shannon form can be found in \cite{PREpow,AIP}. A more systematic analysis of this issue is left for  future studies. In what follows we are focusing on a relation between our results and 
the recently proposed variational definition of the entropy functional \cite{Wang1}. 
\section{Relation to the variational entropic form}
For a probability distribution $f(x)$ of a random variable $x$ a variational entropic form
\begin{equation}
dI=\eta \left[d{\overline x}-{\overline{ dx}}\right]=\eta\int{dxx df(x)}
\label{var}
\end{equation}
has recently been proposed \cite{Wang1} as a measure of uncertainty. Here the overbar again denotes the corresponding averages (we will consider continuous variables)
\begin{equation}
{\overline{x}}=\int{dx x f(x)}
\label{overbar}
\end{equation}
and $\eta$ is a constant of the definition. 
This form has been used for solving an inverse problem, that is finding a functional form of $I$ from a given distribution $f(x)$. It has been found that for an exponential distribution the functional $I$ (found from (\ref{var})) is of the Shannon form
\begin{equation}
I=-\int{dx f(x)\ln{f(x)}}
\label{shannon}
\end{equation} 
and for a $q-$exponential distribution the functional $I$ is of the Tsallis (earlier known as Havrda-Charvat) form
\begin{equation}
I=-\frac{1}{1-q}\int{dx [f(x)-f^q(x)]}
\label{tsallis}
\end{equation}
Surprisingly, this is coherent with the results of the direct problem. Namely, maximizing
the Shannon and Tsallis functionals with the constraint (\ref{overbar}) gives, respectively, an exponential and $q-$exponential distributions.  The author claims \cite{Wang1} that "this is not an ordinary and fortuitous mutual invertibility, since the probability and the entropy are not reciprocal functions". 

In what follows we are going to demonstrate that there is a deep link between the maximum entropy approach and the variational definition (\ref{var}). 
It should be noted that before starting the discussion it is necessary to specify
the meaning of the distribution variation $d f(x)$ in (\ref{var}). In the spirit of the inverse problem mentioned above the variation means a change of the functional form of $f(x)$. In the context of our study it is a variation of the conditional distribution shape $f(x|\mu)$ with the parameter $\mu$.
Starting from eq.~(\ref{vvar0}) and
taking into account that $dA(\mu)=[\partial A(\mu)/\partial \mu]d\mu$ we recover the variational
form quite similar to (\ref{var})
\begin{equation}
dH(\mu)=
-\kappa\int{dx \theta(\mu|x)df(x|\mu)}
\label{vvar}
\end{equation}
In contrast to (\ref{var}), here $\kappa$ is not an arbitrary constant as $\eta$ in eq.~(\ref{var}). It is the Lagrange multiplier associated to the constraint (\ref{constraint}).
Note that in deriving this equation the maximum entropy condition (\ref{variation}) has been used. Therefore we have a relation between the entropy and the distribution which it maximizes. In other words, we have demonstrated that the definition (\ref{var}) is consistent with the maximum entropy principle for a quite arbitrary entropy form. Thus, starting from (\ref{var}) one can try to recover the entropy functional that is maximized by a given distribution with a given constraint (if such an inverse problem has some practical interest \cite{Abe,Wang1}). In this respect it is not surprising that the functionals found \cite{Wang1,Wang2} from (\ref{var}) with a set of probability distributions coincide with the functionals giving these distribution in the maximum entropy approach. Namely, considering (\ref{vvar}) as a definition of the so-called maximizable entropy $H(\mu)=H[f(x|\mu)]$ and 
following the scheme proposed in \cite{Wang1,Wang2}, for the formally exponential distribution
\begin{equation}
f(x|\mu)=\frac{\exp(-\kappa\theta(\mu|x))}{\int dx \exp(-\kappa\theta(\mu|x))}
\label{expf}
\end{equation}
we obtain the Shannon form for the entropy 
\begin{equation}
H[f(x|\mu)]=-\int dx f(x|\mu)\ln f(x|\mu) + const 
\end{equation}   
Nevertheless, if the relevant variable is $x$, then the actual form of the distribution (\ref{expf}) depends on the form of the constrained function $\theta(\mu|x)$. For instance, as we have demonstrated \cite{PREpow,AIP}, if
$\theta(\mu|x)$ is of logarithmic form, then (\ref{expf}) transforms into a $\Gamma$-distribution. Physically this corresponds to a non-equilibrium stationary state whose thermodynamic entropy is held at a given distance from the maximum that corresponds to the equilibrium.  For $\theta(\mu|x)=g(\mu)x^{\alpha}$ we obtain from (\ref{expf}) a stretched exponential distribution. 
This means that we can associate exponential distributions to the Shannon entropy only in the case of linear constraint (\ref{overbar}), i.e. $\theta(\mu|x)=g(\mu)x$.      
   
\section{Conclusion}
The maximum entropy procedure operating with a quite general entropic functional is considered. This allows us to make a link between the maximum entropy approach and the variational formulation \cite{Wang1,Wang2} of the inverse problem. It is shown that in general one cannot uniquely associate a given distribution to a given entropic functional because a form of the constrained function  ($\theta(\mu|x)$ in our case) also contributes
to the actual form of the distribution. On the other hand, we are not free to change the constraint in {\it ad hoc} manner without some physical (or logical) reasons because it encodes our partial information $\theta(\mu)$ into what we are supposed to know ($\theta(\mu|x)$) and what we are searching for ($f(x|\mu)$). 
On the author's opinion a more constructive way is to find a well axiomatically established ("universal") entropy functional suitable for all applications. Then the strategy would be quite clear. That is, if a given distribution cannot be obtained through the maximization of the entropy with a given constraint, then the information (or physics) encoded in the constraint is not compatible with the observed behavior. In such a situation one has to search for another scenario (represented by $\theta(\mu|x)$) of the processes running in the system of interest. 

 


\end{document}